\documentclass[12pt]{amsart}
\usepackage{geometry}                
\usepackage{graphicx,color,bm,float,soul}
\usepackage{amsmath,amssymb,amsthm,amsfonts} 
\usepackage{bbm}
\usepackage{times}
\usepackage{cite}
\usepackage{enumerate}

\newcommand{\cut}[1]{}
\newcommand{\ignore}[1]{}
 \newcommand{\beq}{\begin{equation}}
  \newcommand{\eeq}{\end{equation}}
  \newcommand{\beql}[1]{\begin{equation}\label{eq:#1}}

  \newcommand{\beqa}{\begin{eqnarray}}
  \newcommand{\eeqa}{\end{eqnarray}}
  \newcommand{\beqas}{\begin{eqnarray*}}
  \newcommand{\eeqas}{\end{eqnarray*}}
    \newcommand{\bal}{\begin{align}}
  \newcommand{\eal}{\end{align}}
  \newcommand{\bals}{\begin{align*}}
  \newcommand{\eals}{\end{align*}}
  \newtheorem{Theorem}{Theorem}[section]

  \newcommand{\Tr}{{\rm Tr}}

\newcommand{\bmat}{\left[\begin{array}{rr}}
\newcommand{\emat}{\end{array}\right]}
\newcommand{\bvec}{\left[\begin{array}{r}}
\newcommand{\evec}{\end{array}\right]}
\newcommand{\btmat}{\left[\begin{array}{rrr}}
\newcommand{\etmat}{\end{array}\right]}

  \newcommand{\cH}{{\mathcal H}}
  \newcommand{\cI}{{\mathcal I}}
  \newcommand{\cK}{{\mathcal K}}

\newcommand{\benum}{\begin{enumerate}[{\rm (i)}]\itemsep=0in}
\newcommand{\eenum}{\end{enumerate}}

\newcommand{\bProof}{\begin{proof}}
\newcommand{\eProof}{\end{proof}}
\newcommand{\bTheorem}{\begin{Theorem}}
\newcommand{\eTheorem}{\end{Theorem}}
\newcommand{\deq}[1]{\begin{align}#1\end{align}}

\title[Bohr's contextuality]
{Formalization of Bohr's contextuality within theory of open quantum systems}
\author[A. Khrennikov]{Andrei Khrennikov\\ 
Linnaeus University, International Center for Mathematical Modeling\\  in Physics and Cognitive Sciences
 V\"axj\"o, SE-351 95, Sweden}
\date{}                     
\begin{document}

\begin{abstract} In quantum physics, the notion of contextuality has a variety of interpretations which are typically associated with the names of their inventors, say Bohr, Bell, Kochen and Specker, and recently Dzhafarov. In fact, Bohr was the first who pointed to contextuality of quantum measurements as a part of formulation of his principle of complementarity. (Instead of  ``contextuality'', he considered  dependence on ``experimental conditions.'') Unfortunately, the contextuality counterpart of the complementarity  principle was overshadowed by 
the issue of incompatibility of observables. And the interest for contextuality of quantum measurements rose again only in connection with the Bell inequality. The original Bohr's contextuality, as contextuality of each quantum measurement, was practically forgotten. 
It was highlighted in the works of the author of this paper, with applications both to physics and cognition. In this note, 
the theory of open quantum systems is applied to formalization of Bohr's contextuality within the  
the scheme of indirect measurements. This scheme  is widely used in quantum information theory and it leads to the theory of quantum instruments (Davis-Lewis-Ozawa). In this scheme, Bohr's viewpoint  on contextuality of quantum measurements is represented in the formal mathematical framework. 
\end{abstract}
\maketitle

keywords: Bohr-contextuality,  indirect measurement scheme, open quantum systems, measurement apparatus, state of system, 
state of apparatus, interaction between system and apparatus, quantum instruments 

\section{Introduction}

Nowadays contextuality defined via violation of  the Bell-type inequalities \cite{Bell1,Bell2} or generally noncontextuality 
inequalities \cite{AD0,AD} is a hot topic in quantum physics. Although this notion is well-formalized  in the mathematical 
framework, its physical meaning (beyond nonlocality) is unclear. The same can be said about Kochen-Specker contextuality \cite{KSP}. The main problem is how to {\it separate contextuality from incompatibility}. For quantum observables, incompatibility is the necessary condition of violation of noncontextuality inequalities \cite{IJTP}. Moreover, for the CHSH-inequality (and its noncontextual inequality counterpart) incompatibility is also the sufficient condition of its violation \cite{NL1,IJTP}, at least for observables having the tensor product structure.
\footnote{In \cite{IJTP}, I tried to find intrinsic contextual component in the CHSH-framework - beyond the tensor structured observables. 
It was shown that generally contextuality without incompatibility may have some physical content. A mathematical
constraint extracting the (Bell-)contextuality component from incompatibility was found.
However, its physical meaning is the subject of further studies.} So, the problem whether Bell-contextuality is reduced to 
incompatibility or not is still open. Unfortunately, the Bell-contextuality community simply ignores this complex problem and 
proceed further with mathematical studies.\footnote{It has to be stressed that this paper is devoted to physics. Outside of physics, 
noncontextual inequalities can be violated even for compatible observables. The best example is the Suppes-Zanotti inequality \cite{SZ} 
for three compatible observables. See \cite{BC1,BC2,BC4,BC6} for violation of Bell-type inequalities for observables in cognitive experiments. However, for such applications these inequalities have to be modified to take into account the presence of signaling. The latter is typical for 
experimental data in cognitive and psychological studies. Such inequalities are derived in the novel approach known as contextuality per default \cite{CPD}.}

However, at the very beginning of quantum mechanics Bohr considered a notion of contextuality having the clear physical meaning
\cite{BR0}:

\medskip

{\it ``Strictly speaking, the mathematical formalism of quantum mechanics
and electrodynamics merely offers rules of calculation for the deduction of
expectations pertaining to observations obtained under well-defined experimental 
conditions specified by classical physical concepts.''}

\medskip

Instead of  ``contextuality'', he considered  dependence on ``experimental conditions.''
\footnote{We remark that neither Bell nor Kochen and Specker  operated with the notion of contextuality. 
It was invented later (see  \cite{KHB3} for details).} But original Bohr's contextuality was completely forgotten\footnote{
It was highlighted only in the works of the author of this paper, with applications both to physics \cite{KC3}-\cite{Google} 
and cognition \cite{QL1}.} and  Bell, Kochen-Specker and their followers considered a variety of its ``derivatives'' related to the very special case of joint measurement of pairs of compatible observables. Heuristically this sort of contextuality is represented as following: 
 
{\bf Joint-measurement contextuality:} If $A, B, C$ are three quantum observables, such that $A$ is compatible with $B$ and  $C,$ a measurement of $A$ might give different result depending upon whether $A$ is measured  with $B$ or with $C.$

This formulation is based on counterfactual argument and from my viewpoint it cannot be tested experimentally, so it has no relation to physics.\footnote{We remark Svozil \cite{KS1,KS2,KS3}  and Griffiths  \cite{GGG} claim that they elaborated experimental tests for  
joint-measurement contextuality.}

Bohr-contextuality is experimentally tested through incompatibility and theoretically it is formulated in terms of commutators.
The basic test is based on the Heisenberg uncertainty relation in its general form of the Schr\"odinger-Robertson inequality. 
structure. Bell-contextuality can be tested experimentally in experiments by demonstration of violation of various Bell-type inequalities.

Since Bohr's formulation of the complementarity principle  including the contextuality counterpart, quantum measurement theory 
was  essentially developed. The most powerful quantum measurement formalism is based on the theory of 
{\it open quantum systems} \cite{ING,ING1}  desribing  a system $S$ interacting  with the surrounding environment $E.$ In the particular application of this theory to the description of measurements, the role of environment is played by a measurement apparatus $M$ used to measure some observable $A$ on $S.$ One of the basic mathematical frameworks for modeling  this situation is {\it the scheme of indirect measurements} \cite{Oza84}. The outcomes of $A$ are represented as outcomes of apparatus' pointer $M_A.$  

The process of measurement is described as the interaction between $S$ and  $M.$ This interaction generates 
the dynamics of the state of the compound system $S+M.$ The state's evolution is unitary; for pure states, it is described        by 
the Schr\"odinger equation (for mixed states given by density operators, by the von Neumann equation). Finally, the probability 
distribution for pointer's outcomes is extracted from the compound state with the trace-operation.

The scheme of indirect measurements is closely coupled to the theory of quantum instruments \cite{Dav76,Oza84,O3},  but in this paper we shall not discuss this issue.

We start the paper with structuring Bohr's complementarity principle \cite{BR0,PL2} and highlighting its contextuality-component (section \ref{BC}). Then we present the indirect measurement scheme (section \ref{IND}) and finally (section \ref{BCIND}) formalize  Bohr's contextuality in this framework. 

\section{Contextuality component of Bohr's principle of   complementarity}
\label{BC}

Here we follow my previous works devoted to Bohr's principle of complementarity and its contextual component \cite{KHR_CONT,KHR_BB,NL1,Google}.
We start with the well known citate of Bohr (\cite{BR0}, v. 2, p. 40-41): 

\medskip

``This crucial point ...  implies {\it the impossibility of any sharp separation between the behaviour of atomic objects and the interaction with the measuring instruments which serve to define the conditions under which the phenomena appear.} In fact, the individuality of the typical quantum effects finds its proper expression in the circumstance that any attempt of subdividing the phenomena will demand a change in the experimental arrangement introducing new possibilities of interaction between objects and measuring instruments which in principle cannot be controlled. Consequently, evidence obtained under different experimental conditions cannot be comprehended within a single picture, but must be regarded as complementary in the sense that only the totality of the phenomena exhausts the possible information about the objects.''
     
\medskip

The contextual component of this statement can be formulated as 

 \medskip

{\bf Principle 1 (Contextuality)} {\it  The output of any quantum observable is indivisibly composed of the contributions of the system and the measurement apparatus.}

\medskip

There is no reason to expect that all experimental contexts
can be combined and all observables can be measured jointly. Hence, 
incompatible observables (complementary experimental contexts) may exist. Moreover, they should exist, otherwise the 
contextuality principle would have the empty content.  Really, if all experimental
contexts can be combined into single context ${\it C}$ and all observables can be jointly measured in this context, then the outputs of such joint measurements can be assigned directly to a system. To be more careful, we have to say: ``assigned to a system and context ${\it- C}''.$ But, the latter can be omitted, since this is the same context for all observables. This reasoning implies:         

\medskip

{\bf Principle 2 (Incompatibility)} {\it There exist incompatible observables (complementary experimental contexts).}

\medskip

Since both principles, contextuality and ncompatibility, are so closely interrelated, it is natural to unify them into the single
principle, {\bf Contextuality-Incompatibility  principle.} This is my understanding of the Bohr's Complementarity principle.

\section{Basics of the quantum formalism}
\label{BQF}

In quantum theory, it is postulated that every quantum system $S$ corresponds to a complex 
Hilbert space $\cH;$ denote the scalar product of two vectors by the symbol $\langle \psi_1\vert \psi_2\rangle.$
Throughout the present paper, we assume $\cH$ is {\it finite dimensional.}
States of the quantum system $S$ are represented by density operators acting in $\cH.$  
Denote this state space by the symbol  $\bf{S}(\cH)$. Observables are  represented by Hermitian operators in $\cH.$ These are just {\it symbolic expressions of phsyical observables}, say the position, momentum, or energy.  Each Hermitian operator 
$ A$ can be represented as 
\begin{equation}
\label{HH}
 A = \sum_x x E^{A}(x),
\end{equation}
where  $x$ labels the eigenvalues and  $E^{A}(x)$ is the spectral projection of the observable $A$ corresponding to 
the eigenvalue $x$. 

The operator $A$ can be considered as the compact mathematical representation for probabilities 
of outcomes of the physical observable. These probabilities are given by {\it the Born rule} that states
 if an observable $A$ is measured in a state $\rho,$ then
the probability distribution $\Pr\{A=x\|\rho\}$ of the outcome of the measurement is given
by
\deq{\label{eq:Born}
\Pr\{A=x\|\rho\}=\Tr[E^{A}(x)\rho] = \Tr[E^{A}(x)\rho E^{A}(x)].
}

\section{Indirect measurement scheme: apparatus with meter interacting with a system}
\label{IND}
 
The scheme of indirect measurements represents the framework which was  emphasized by Bohr, by him the outcomes of quantum measurements are created in the complex process of the interaction of a system $S$ 
with a measurement apparatus $M.$ The latter is combined of a complex physical device interacting with $S$ and a pointer 
showing the outcomes of measurements; for example, it can be the ``spin up or spin down'' arrow. 
The system $S$ by itself is not approachable by the observer who 
can see only the pointer of $M.$ Then the observer associates pointer's outputs with the values of 
measured observable $A$ for the system $S.$ 

Can the outputs of the pointer be associated with the ``intrinsic properties'' of
$S$  or not? This is one of the main questions of disturbing the quantum foundations during the last 100 years.

The indirect measurement scheme can be represented as the block of following interrelated components:
\begin{itemize}
\item the states of the systems $S$ and the apparatus $M;$ they are represented in complex Hilbert spaces 
$\cH$ and $\cK,$ respectively;    
\item the unitary operator $U$ representing the interaction-dynamics for the compound system $S+M;$ 
\item the meter observable $M_A$  giving outputs of the pointer of the apparatus $M.$ 
\end{itemize}

In the indirect measurement scheme, it is assumed that the compound system  $S+M$ is isolated. 
The dynamics of pure states of  the compound system is described by the Schr\"odinger equation: 
\begin{equation}
\label{SCHET}
i\frac{d}{dt} \vert \Psi\rangle(t)=     H  \vert \Psi\rangle(t), \; \vert \Psi\rangle(0)= \vert \Psi\rangle_0,
\end{equation} 
where $ H $ is it Hamiltonian (generator of evolution) of  $S+M$. 
The state $\vert \Psi\rangle(t)$ evolves as 
$$
\vert \Psi\rangle(t)=    U(t)\vert \Psi\rangle_0,
$$ 
where $U(t)$ is the unitary operator represented as  
$$   
U(t)= e^{-i t     H }.
$$ Hamiltonian (evolution-generator) describing information interactions has the form 
$$
     H  =   H_S\otimes I +    I \otimes H_{M} +    H_{S,M},
$$ 
where $H_S: \cH  \to \cH ,    H_{M}: \cK  \to \cK $ are Hamiltonians of $S$ and $M,$ 
respectively,  and $   H_{S,M} \in \cH  \otimes \cK  \to \cH  \otimes \cK $ is  Hamiltonian of interaction between 
systems $S$ and $M.$
 The Schr\"odinger equation implies that evolution of the density operator $R(t)$ of the system $S+ M$ is described by the von Neumann equation: 
\begin{equation}
\label{VNEE}
 \frac{d   R}{d t}(t)= - i [  H ,R(t)],\;    R(0)=    R_0.
\end{equation}
However, the state $   R(t)$ is too complex to be handled consistently: the apparatus includes many degrees of freedom.

Suppose that we want to measure an observable on the system $S$ which is represented by Hemitian operator $A,$ 
acting in system's state space $\cH.$  The {\em indirect measurement model} for measurement of the $A$-observable  was introduced by Ozawa in \cite{Oza84} as a ``(general) measuring process''; this is a quadruple 
$$
(\cK, \sigma, U, M_A)
$$ 
consisting of a Hilbert space $\cK,$   a density operator $\sigma \in \bf{S}(\cK),$ a unitary operator $U$ 
on the tensor product of the state spaces of $S$ and $M,$ $U: \cH\otimes\cK \to \cH\otimes\cK,$ and a Hermitian operator $M_A$ on $\cK$.  

Here  $\cK$ represents the states of the apparatus  $M$,  $U$ describes the time-evolution of  system $S+ M$,  $\sigma$ describes the initial state of the apparatus $M$ before the start of measurement, and the Hermitian operator $M_A$ is the meter observable of the apparatus $M$ (say the pointer of $M).$ This operator represents indirectly outcomes of an observable $A$ for the system $S.$ 

The  probability distribution $\Pr\{A =x\|\rho\}$ in the system state $\rho \in \bf{S}(\cH)$
is given by
\beq
\label{MA1}
\Pr\{A =x\|\rho\}=\Tr[(I\otimes E^{M_A}(x))U(\rho\otimes\sigma)U^{\star}],
\eeq 
where $E^{M_A}(x)$ is  the spectral projection of $M_A$ for the eigenvalue $x.$
We reall that operator $M_A$ is  Hermitian. In the finite dimensional case, it can 
be represented in the form:
\beq
\label{MA2}
M_A= \sum_k x_k E^{M_A}(x_k),
\eeq 
where $(x_k)$ is the set of its eigenvalues and $E^{M_A}(x_k)$ is the projector 
on the subspace of eigenvectors corresponding to eigenvalue $x_k.$

The change of the state $\rho$ of the system $S$  caused by the measurement for the outcome $A =x$ 
is represented with the aid of the map $\cI_A(x)$ in the space of density operators defined as 
\beq
\label{MA1A}
\cI_A(x)\rho =\Tr_{\cK}[(I\otimes E^{M_A}(x))U(\rho\otimes\sigma)U^{\star}],
\eeq 
where $\Tr_{\cK}$ is the partial trace over $\cK.$ 
The map  $x \mapsto \cI_A(x)$  is a quantum instrument. 
We remark that conversely any quantum instrument can be represented via the indirect 
measurement model (see Ozawa \cite{Oza84}).  

\section{Bohr's contextuality from the indirect measurement scheme}
\label{BCIND}

We take the basic part of the aforementioned citate of Bohr:  ``.. the impossibility of any sharp separation between the behaviour of atomic objects and the interaction with the measuring instruments'' and establish correspondence with the  indirect measurement scheme.
\begin{itemize}
\item ``atomic object'' - state $\rho;$
\item ``measuring instrument'' - state $\sigma;$
\item ``interaction'' - unitary operator $U.$
\end{itemize}

In this framework, the triple ${\it C}= (\rho, \sigma, U)$ represents the complex of the ``experimental conditions'', the 
context of measurement. In this framework, the relation between contextuality (in Bohr's sense) and incompatibility is completely 
clear: incompatibility is so to say the ``derivative'' of contextuality. There is no reason to expect that any pair of contexts,
${\it C}_1= (\rho_1, \sigma_1, U_1)$  and ${\it C}_2= (\rho_2, \sigma_2, U_2)$ can be unified in the joint measurement scheme, 
even if $\rho_1=\rho_2= \rho.$  

\section{Concluding remarks}

Recently Bell and Kochen-Specker contextualities attracted a lot of attention; interesting studies were performed and 
they definitely made the mathematical  structures of  these contextualities more complete. However, the physical meaning
of these contextualities is still not clear. The common way to connect Bell-contextuality with physics is to refer to nonlocality and 
spooky action at the distance. However, many authors think that the nonlocality issue is not crucial  (\cite{NL2}--\cite{MM}). Another way  to connect Bell-contextuality with physics 
is to couple it to incompatibility \cite{KHR_BB,NL1,NL2,NL3} and generally to the Bohr's complementarity principle. As was mentioned in introduction, for 
the CHSH-scheme (and ``natural observables'') Bell-contextuality is identical to incompatibility. This is really the alarming signal 
for those who use violation of Bell-type inequalities to quantify contextuality.  (Of course, interrelation incompatibility-contextuality
has to be investigated in more details, especially for noncontextual inequalities with $n \geq 5$ observables.)  

At the same time the original Bohr's viewpoint on contextuality of quantum measurements and its connection with incompatibility is practically forgotten. In this paper,  Bohr's viewpoint was refreshed (section \ref{BC}) in the form of Principle 1 (contextuality) and
Principle 2 (principle 2). Bohr-contextuality was formalized within the scheme of indirect measurements.

\section*{Acknowledgments} 

The author would like to thank Masanao Ozawa for discussions on the indirect measurement scheme.

\end{document}